\documentclass[preprint]{elsarticle}
\usepackage{color}
\usepackage[bookmarks,bookmarksnumbered]{hyperref}
\hypersetup{
    colorlinks = true,
    linkcolor = blue,
    anchorcolor =red,
    citecolor = blue,
    filecolor = red,
    urlcolor = red,
    pdfauthor=author
}
\usepackage{todonotes}
\usepackage{subfigure}
\usepackage{ulem}
\usepackage{amssymb}
\usepackage{booktabs}
\usepackage{chemformula}
\usepackage{gensymb}
\usepackage{tikz}
\usepackage{graphicx}
\usepackage{physics}
\usepackage{dcolumn}
\usepackage{bm}
\usepackage{xcolor}
\setchemformula{radical-radius=1pt}
\RequirePackage{xspace}
\DeclareFontFamily{OT1}{pzc}{}
\DeclareFontShape{OT1}{pzc}{m}{it}{<-> s * [1.200] pzcmi7t}{}
\DeclareMathAlphabet{\mathpzc}{OT1}{pzc}{m}{it}
\usepackage{amsfonts}

\journal{Nuclear Instruments and Methods A}

\usepackage{multirow}
\newcommand{\doserate}{\mathcal{R}}

\usetikzlibrary{decorations.pathmorphing}
\pgfdeclaredecoration{complete sines}{initial}{
  \state{initial}[
    width = +0pt ,
    next state = sine,
    persistent precomputation = {
      \pgfmathsetmacro\matchinglength{
        \pgfdecoratedinputsegmentlength /
        int(\pgfdecoratedinputsegmentlength/\pgfdecorationsegmentlength)
      }
      \setlength{\pgfdecorationsegmentlength}{\matchinglength pt}
    }
  ]{}
  \state{sine}[width=\pgfdecorationsegmentlength]{
    \pgfpathsine{
      \pgfpoint
        {0.25\pgfdecorationsegmentlength}
        {0.5\pgfdecorationsegmentamplitude}
      }
    \pgfpathcosine{
      \pgfpoint
        {0.25\pgfdecorationsegmentlength}
        {-0.5\pgfdecorationsegmentamplitude}
      }
    \pgfpathsine{
      \pgfpoint
        {0.25\pgfdecorationsegmentlength}
        {-0.5\pgfdecorationsegmentamplitude}
      }
    \pgfpathcosine{
      \pgfpoint
        {0.25\pgfdecorationsegmentlength}
        {0.5\pgfdecorationsegmentamplitude}
      }
  }
  \state{final}{}
}
\tikzset{
  wave/.style={
    decorate,decoration={
      complete sines,
      segment length=3pt,
      amplitude=2pt
    }
  }
}
\NewChemArrow{w>}{
    \path (cf_arrow_start) -- +(+5pt,0) coordinate (xxx1) ; 
    \path (cf_arrow_end) -- +(-8pt,0) coordinate (xxx2) ; 
    \draw[-] (cf_arrow_start) -- (xxx1) ;
    \draw[chemarrow,wave] (xxx1) -- (xxx2) ;
    \draw[chemarrow,-cf] (xxx2) -- (cf_arrow_end);
}

\begin{document}

\begin{frontmatter}

\title{Effects of oxygen on the optical properties of\\phenyl-based scintillators during irradiation and recovery}

\author[umd]{C.~Papageorgakis\corref{cor}}
\ead{cpapag@umd.edu}
\author[umd]{M.Y.~Aamir}
\author[umd]{A.~Belloni}
\author[umd]{T.K.~Edberg}
\author[umd]{S.C.~Eno}
\author[umd]{B.~Kronheim}
\author[umd]{C.~Palmer}

\affiliation[umd]{ 
    organization={Dept. Physics, U. Maryland},
    city={College Park},
    state={MD},
    country={USA}
}

\begin{abstract}
Plastic scintillators are a versatile and inexpensive option for particle detection, which is why the largest particle physics experiments, CMS and ATLAS, use them extensively in their calorimeters. One of their challenging aspects, however, is their relatively low radiation hardness, which might be inadequate for very high luminosity future projects like the FCC-hh. In this study, results on the effects of ionizing radiation on the optical properties of plastic scintillator samples are presented. The samples are made from two different matrix materials, polystyrene and polyvinyltoluene, and have been irradiated at dose rates ranging from $2.2\,$Gy/h up to $3.4\,$kGy/h at room temperature. An internal boundary that separates two regions of different indices of refraction is visible in the samples depending on the dose rate, and it is compatible with the expected oxygen penetration depth during irradiation. The dose rate dependence of the oxygen penetration depth for the two matrix materials suggests that the oxygen penetration coefficient differs for PS and PVT. The values of the refractive index for the internal regions are elevated compared to those of the outer regions, which are compatible with the indices of unirradiated samples.
\end{abstract}


\end{frontmatter}

\section{\label{sec:intro}Introduction}
\subsection{\label{subsec:plasticScint}Plastic scintillators}

The use of plastic scintillators in high-energy physics experiments has been widespread both in past and present experiments. Their low cost, ease of production, and versatility make them attractive as the active element for a broad range of particle detectors, with calorimeters being the main beneficiaries. Looking at past experiments, the Collider Detector at Fermilab (CDF) used a hadronic calorimeter made out of alternating sheets of plastic scintillator and steel \cite{CDF:1987vot}, and the DØ experiment had its preshower detector \cite{SMIRNOV200994} and outer tracker made out of plastic scintillator panels and fibers \cite{LINCOLN1996424}, respectively. Moving to the present, both the A Toroidal LHC Apparatus (ATLAS) and the Compact Muon Solenoid (CMS) experiments, located at the Large Hadron Collider (LHC) at CERN, make use of plastic scintillator. For ATLAS, the Tile Calorimeter (TileCal) consists of steel modules with slots for plastic scintillating tiles equipped with wavelength-shifting fibers (WLS) \cite{HERNANDEZ201383}. For CMS, the Hadronic Calorimeter (HCAL) uses alternating layers of brass and plastic scintillating tiles with WLS fibers embedded for readout \cite{Mans:1481837}. The Phase 2 upgrades for CMS, which are scheduled to be installed in 2026, include the new High Granularity Calorimeter (HGCAL) that is going to use plastic scintillator tiles instrumented with Silicon Photomultipliers (SiPMs) for the low radiation regions of its hadronic part \cite{Ochando:2311394}. Finally, future experiments are considering the use of plastic scintillators, e.g., the IDEA detector at the Future Circular Collider for electrons and positrons (FCC-ee), which is contemplating the use of plastic scintillating fibers in its dual readout calorimeter \cite{Antonello_2020}.

Plastic scintillators consist of a substrate material that is typically doped with a primary fluor at a concentration of a few percent and a secondary fluor at a lower concentration. The substrate, which is usually either polystyrene (PS) or polyvinyltoluene (PVT), acts as the main receptacle for the ionizing particle's energy. This energy is then transferred to the primary dopant either through the non-radiative F\"orster mechanism \cite{forster} for higher dopant concentrations ($\gtrsim1\%$) \cite{birks} or through radiative transfer for lower concentrations. The primary dopant emits light of wavelengths that overlap with the secondary dopant absorption spectrum. The secondary dopant receives the energy from the primary through radiative transfer and then emits light in the visible part of the spectrum.

One crucial aspect of detector design for HEP experiments is radiation tolerance. Ionizing radiation is capable of breaking the polymer molecules and creating radicals, which are molecules with one or more unpaired valence electrons. The radicals are known to strongly absorb light of shorter wavelengths and they tend to react easily with other radicals, in a group of processes called radical recombination that includes radical crosslinking, or with the oxygen dissolved in the material, thus getting oxidized. Therefore, radiation damage can be defined as the collective effect of all those chemical processes on the scintillation mechanism and the attenuation length of the material under study.

The collisions facilitated by particle accelerators are capable of creating high particle fluxes, with the regions closer to the interaction point and the endcaps being affected disproportionately. Modern experiments are routinely experiencing total absorbed doses $\sim1\,$MGy for vertex detectors \cite{Adam_2021} close to the interaction point, and $>1\,$kGy at the calorimeter endcaps, where plastic scintillator is often found.  Endcap doses are accumulated at very low dose rates ranging between $10^{-3}$ and $1\,$Gy/h. Specifically, at CMS, during the $50\,$fb$^{-1}$ of integrated luminosity gathered at 13 TeV in 2017, the Hadronic Endcap (HE) tiles of HCAL received doses up to a few kGy \cite{Sirunyan_2020}. The High Luminosity (HL) era at the LHC is going to bring new challenges with respect to radiation tolerance. The anticipated high pileup of collisions are expected to increase further the expected total absorbed doses up to $\sim3\,$kGy for the plastic scintillator that is going to be installed at the CMS HGCAL~\cite{hgcaltdr}. 

Numerous studies of radiation-induced damage in scintillators exist in literature~\cite{Jivan_2015,Pedro_2019,sauli,dubna,Biagtan1996125,34504,Wick1991472,289295,173180,173178,Giokaris1993315,1748-0221-11-10-T10004,gillen}. However, most studies use data collected by high dose rate irradiations or are focused on scintillators read out with WLS fibers, and therefore the results describe the combined damage. In more recent works, lower dose rates have been studied in more detail, and the importance of the dose rate in radiation damage has been further established \cite{PAPAGEORGAKIS2022167445}. In addition, the presence (or absence) of oxygen during irradiation and recovery of the scintillators affects radiation damage significantly \cite{cloughquant,clough2,GILLEN1995149,HORSTMANN1993395}. This work will present studies relating oxygen diffusion in polymers with some of the radiation-induced optical properties of scintillators.

\section{\label{sec:oxygenDiffusionTheory}Oxygen diffusion theory}

This section presents an overview of the oxygen diffusion theory in polymers, followed by a discussion of two models that incorporate the effect of oxidation on oxygen diffusion. These two models follow two different approaches, the first having been developed for oxidation induced by ionizing radiation and the second for photo-oxidation.

\subsection{\label{subsec:diffusionInPolymers}Oxygen diffusion in polymers}

The diffusion of a gas into a solid or liquid material is described by Fick's two laws of diffusion \cite{doi:10.1002/andp.18551700105, doi:10.1080/14786445508641925}. The first Fick's law is for the diffusion flux $J$
\begin{equation}\label{eq:ficks1st}
    J = -D\frac{dC}{dx},
\end{equation}
where $D$ is the diffusion coefficient, and $C$ is the concentration. The second Fick's law is the diffusion equation
\begin{equation}\label{eq:ficks2nd}
    \frac{\partial C}{\partial t}=D\frac{\partial^2 C}{\partial x^2}.
\end{equation}

The dissolved amount of a gas in a solid or liquid at equilibrium is described by Henry's law \cite{doi:10.1098/rstl.1803.0004}
\begin{equation}\label{eq:henryslaw}
    C = Sp,
\end{equation}
where $p$ is the partial pressure of the gas at the surface of the material, and  $S$ is the solubility (or Henry) coefficient. The polymers used in scintillators are usually amorphous and have a low degree of crystallinity. For these polymers, which are commonly characterized as glassy, two solubilities can be defined, one for diffusion in the material, following Henry's law, and one for diffusion in the voids in the material that is described by Langmuir isotherms \cite{vieth,neogi1996diffusion}.

For a thin membrane, Fick's second law can be rewritten using Henry's law as
\begin{equation}\label{eq:ficks2ndPermeability}
    J=-DS\frac{\Delta p}{\delta}=-\bar{P}\frac{\Delta p}{\delta},
\end{equation}
where $\Delta p$ is the pressure difference between the two sides of the membrane, $\delta$ is its thickness, and $\bar{P}$ is the permeability.

The quantities described in the previous paragraphs have temperature dependencies. Specifically, the solubility $S$ obeys a van 't Hoff equation \cite{Stern2007}
\begin{equation}\label{eq:vanthoff}
    S = S_0 e^{-\frac{\Delta H_S}{RT}},
\end{equation}
where $\Delta H_S$ is the enthalpy change, $R$ is the universal gas constant, and T is the temperature in degrees Kelvin. The diffusivity and the permeability are known to be described by the Arrhenius equation \cite{vieth,Stern2007}
\begin{equation}\label{eq:arrheniusPermeability}
    \bar{P} = \bar{P}_0 e^{-\frac{E_p}{RT}},
\end{equation}
where $E_p$ is the apparent energy of activation of the permeation process, and
\begin{equation}\label{eq:arrheniusDiffusion}
    D = D_0 e^{-\frac{E_d}{RT}},
\end{equation}
where $E_d$ is the activation barrier for diffusion.

\begin{figure}[bt]
    \centering
    \includegraphics[width=0.7\textwidth]{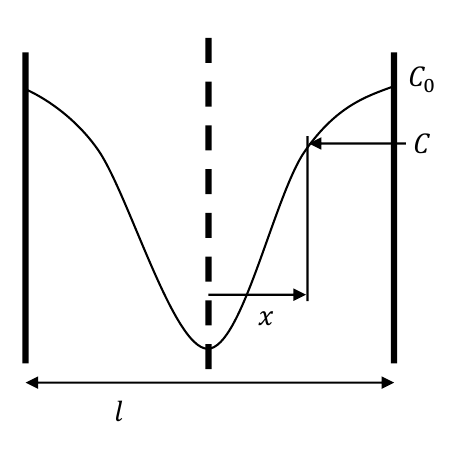}
    \caption{A 1-dimensional description of oxygen diffusion in the polymer. The thickness is $l$, $C_0$ and $C(x,t)$ the oxygen concentrations at the surface and the interior of the polymer, and $x$ the distance from the center.}
    \label{fig:model}
\end{figure}

In the following section, the polymer model that was introduced by Seguchi \textit{et al.} in \cite{seguchi1977radiation,seguchi} is studied. The model, a schematic of which is shown in Fig.~\ref{fig:model},
describes a polymer film of thickness $l$ with oxygen concentration $C(x,t)$ that obeys the following boundary conditions
\begin{enumerate}
    \item $C=C_0$ at $x=\frac{l}{2},-\frac{l}{2}$ for all $t$,
    \item $C=0$ for $t=0$ and $\frac{l}{2}>x>-\frac{l}{2}$,
    \item $\frac{\partial C}{\partial x}=0$ at $x=0$ for all $t$.
\end{enumerate}

Fick's second law of diffusion (Eq.~\ref{eq:ficks2nd}) describes this situation. The solution to this equation is well-known in literature~\cite{seguchi1977radiation,crank1968diffusion}:
\begin{eqnarray}
    \frac{C(x,t)}{C_0}&=&\sum^{\infty}_{n=0}(-1)^n\textrm{erfc}\frac{(2n+1)\frac{l}{2}-x}{2\sqrt{Dt}}+\nonumber\\
    & &\sum^{\infty}_{n=0}(-1)^n\textrm{erfc}\frac{(2n+1)\frac{l}{2}+x}{2\sqrt{Dt}}.
\end{eqnarray}

\subsection{\label{subsec:oxidation}JAERI model for oxidation during irradiation}

The work by Seguchi \textit{et al.} in \cite{seguchi1977radiation,seguchi} describes in detail the chemical reactions involved with oxidation during irradiation and studies the effect of oxygen diffusion at equilibrium for that reaction chain. In this section, these calculations are rederived following the manner of \cite{seguchi1977radiation,seguchi}.

The reactions related to irradiation and oxygen for a polymer \ch{P} can be described by the following formulas that describe the radical formation process
\begin{equation}
     \ch{P w>[ $Y\doserate$ ] P^.},
\end{equation}
where $\ch{P^.}$ is the symbol for the polymer radicals, $Y$ is the specific rate of radical formation, and $\doserate$ is the dose rate; the radical recombination processes, including crosslinking, have the general form
\begin{equation}
    \ch{P^. + P^. ->[ $k_1$ ] P-P},
\end{equation}
where $k_1$ is the rate constant for radical recombination; and the radical oxidation process
\begin{equation}
    \ch{P^. + O2 ->[ $k_2$ ] P-O2},
\end{equation}
where $k_2$ is the rate constant for the oxidation process.

The equation that describes the balance between the polymer radicals and oxygen concentration can be formulated as
\begin{equation}\label{eq:radicals1}
    \frac{d[\ch{P^.}]}{dt} = Y\doserate - k_1 [\ch{P^.}]^2 - k_2 [\ch{P^.}] C(x,t),
\end{equation}
where $[\ch{P^.}]$ is the density of polymer radicals, $C(x,t)$ is the oxygen concentration, and $x$ is the distance from the center of the polymer's cross-section. In steady state, the first time derivative is 0, and Eq.~\ref{eq:radicals1} becomes
\begin{equation}\label{eq:radicals2}
    [\ch{P^.}] = \frac{-k_2 C(x) + \sqrt{(k_2 C(x))^2+4k_1 Y\doserate}}{2k_1}.
\end{equation}

For the oxygen concentration balance, we find
\begin{equation}\label{eq:oxygen}
    \frac{\partial C(x,t)}{\partial t} = D\frac{\partial^2 C(x,t)}{\partial x^2} - k_2 [\ch{P^.}] C(x,t),
\end{equation}
where $D$ is the diffusion coefficient for oxygen in the polymer. The following boundary conditions are assumed
\begin{enumerate}
  \item For $t=0$ and $-\frac{l}{2}\leq x\leq\frac{l}{2}$, $C=C_0$,
  \item At $x=0$ and for all $t$, $\frac{dC}{dx}=0$,
  \item At $x=\pm\frac{l}{2}$ and for all $t$, $C=C_0$.
\end{enumerate}

\subsubsection{If $\frac{4k_1 Y\doserate}{(k_2 C)^2}\ll1$ for all $x$ \label{cond1}}

\begin{figure}[bt]
    \centering
    \includegraphics[width=1\textwidth]{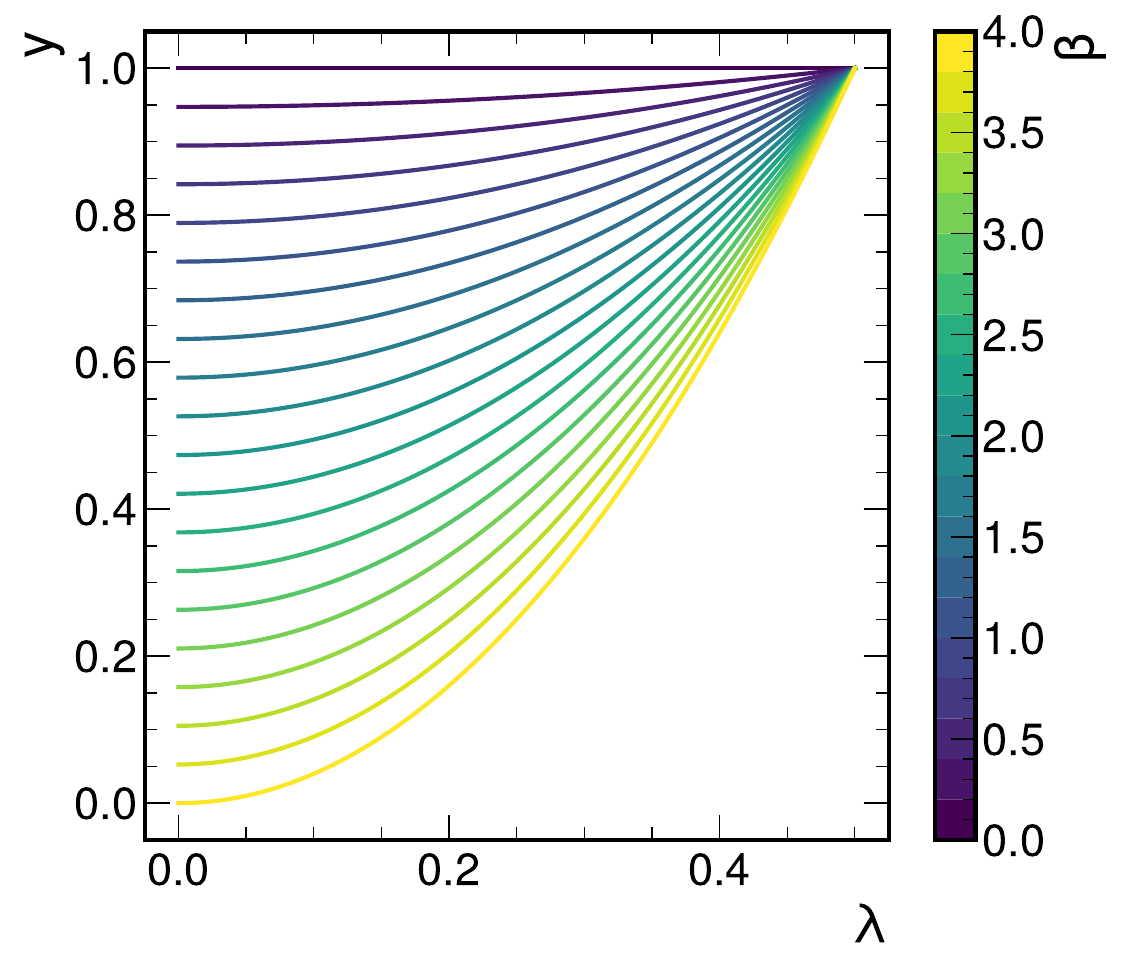}
    \caption{The normalized oxygen concentration $y$ at stationary state under irradiation is plotted for the limiting case $\frac{4k_1 Y\doserate}{(k_2 C)^2}\ll1$ and for a range of values of $\beta$.}
    \label{fig:oxSol}
\end{figure}

When the combined rate of radical creation and crosslinking is much lower than the oxygen bonding rate throughout the sample then the inequality $\frac{4k_1 Y\doserate}{(k_2 C)^2}\ll1$ is true for all $x$, and Eq.~\ref{eq:radicals2} can be simplified by Taylor-expanding the square root to
\begin{equation}
    [\ch{P^.}] = \frac{Y\doserate}{k_2 C(x,t)}.
\end{equation}
By substituting this result into Eq.~\ref{eq:oxygen} we arrive at
\begin{equation}\label{eq:oxygenSimplified}
    \frac{\partial C(x,t)}{\partial t} = D\frac{\partial^2 C(x,t)}{\partial x^2} - Y\doserate.
\end{equation}
For the steady-state case, Eq.~\ref{eq:oxygenSimplified} becomes 
\begin{equation}
    D\frac{\partial^2 C(x)}{\partial x^2} = Y\doserate.
\end{equation}
The solution to this differential equation is 
\begin{equation}\label{eq:solution}
    1 - y = \beta \left(\frac{1}{4} - \lambda^2\right),
\end{equation}
where $y\equiv\frac{C}{C_0}$, $\lambda\equiv\frac{x}{l}$, and
\begin{equation}
    \beta = \frac{Y\doserate l^2}{2DC_0}.
\end{equation}
The steady-state solutions for different $\beta$ are shown in Fig.~\ref{fig:oxSol}
. The code that reproduces this figure can be found in \cite{repository}. For low values of $\beta$, the effect on oxygen concentration is small. The concentration in the center of the sample becomes lower as the value of $\beta$ grows until the limiting case in which $\beta=4$ and $y=0$ for $\lambda=0$. Eq.~\ref{eq:solution} is not a good description for $\beta\gtrsim4$.

If the system is not in steady state, then Eq.~\ref{eq:oxygenSimplified} can only be solved numerically by approximating the derivatives of $C(x,t)$ via a finite differences method.

\subsubsection{\label{cond2}If $\frac{4k_1 Y\doserate}{(k_2 C)^2}\ll1$ is not true for all $x$}
In some cases, oxygen is consumed at rates that cannot be replenished by diffusion, and the concentration can become 0 in the central region of the polymer. This is true when the dose rate is sufficiently high for a given set of the rate constants $k_1$, $k_2$, and $Y$. The differential equation is not analytically solvable in this case but it can be approximated using a finite differences method. In considering the limiting case $C(0, t)=0$, according to Eq.~\ref{eq:solution}, $\beta=4$. Then
\begin{equation}
    l^2=\frac{8DC_0}{Y\doserate}.
\end{equation}
For the oxygen penetration depth $z\equiv l/2$, thus
\begin{equation}\label{eq:depth}
    z^2=\frac{2DC_0}{Y\doserate}.
\end{equation}

\subsubsection{\label{numSol1}Numerical solution}

\begin{figure}[bt]
\centering
\includegraphics[width=1\textwidth]{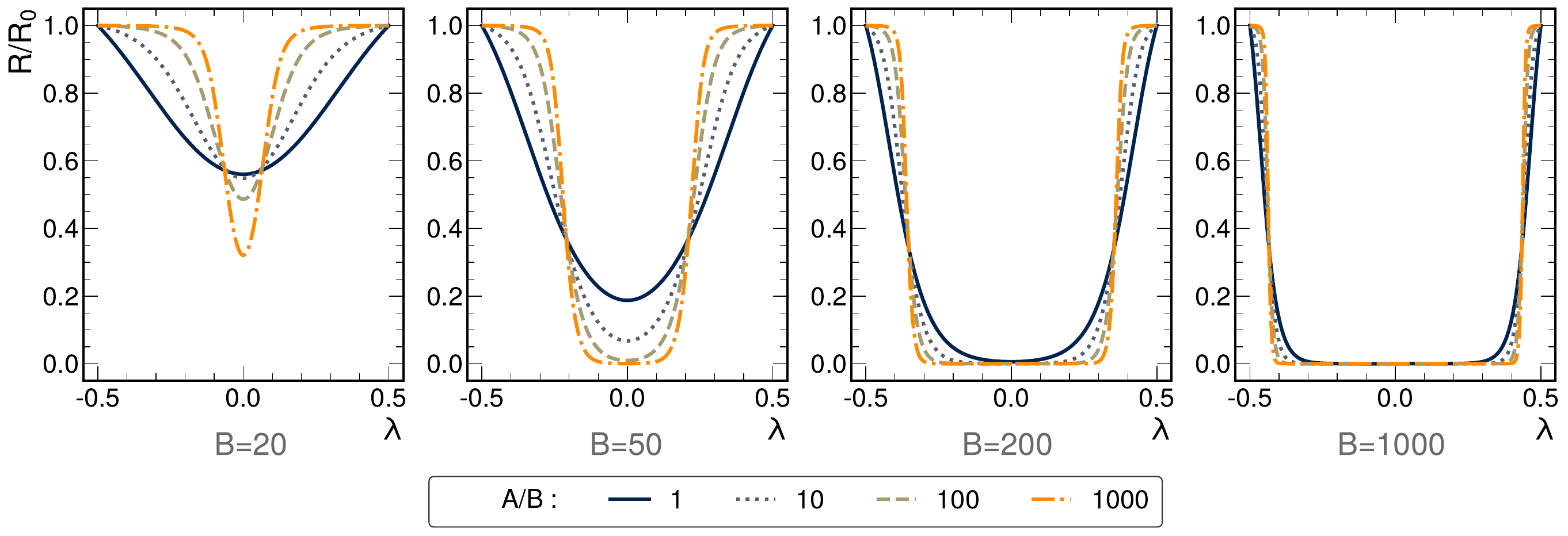}
\caption{\label{fig:wide2}Numerical solutions to Eq.~\ref{eq:numEquation} for various combinations of values for the parameters $A\equiv\frac{C_0k_2^2}{2Dk_1}$ and $B\equiv\frac{2Y\doserate}{DC_0}$. The parameter $A$ is expressed through its ratio over $B$. The normalized reaction rate $R/R_0$ is shown for the entire normalized length $\lambda$. For the code used see Ref.~\cite{repository}.}
\end{figure}

The system of Equations~\ref{eq:radicals2} and \ref{eq:oxygen} can be solved numerically by using \verb|scipy|'s \cite{2020SciPy-NMeth} ODE solver for boundary value problems. The full code can be found in \cite{repository}. The formula to be solved is
\begin{equation} \label{eq:numEquation}
    \frac{d^2y}{d\lambda^2}=Ay(\sqrt{y^2+\frac{B}{A}}-y),
\end{equation}
where $A\equiv\frac{C_0k_2^2}{2Dk_1}$, $B\equiv\frac{2Y\doserate}{DC_0}$. Fig.~\ref{fig:wide2} shows solutions of the normalized reaction rate $R/R_0$ for different combinations of the input parameters $A$ and $B$. Parameter $B$ determines whether an oxygen-depleted region will form in the middle of the sample, as well as how deep in the scintillator the boundary of the region will be. The ratio $A/B$ controls how steep the transition will be. In particle colliders, plastic scintillators are usually placed in relatively low $\doserate$ regions so the curves corresponding to lower values of $B$ and higher values of $A/B$ should be more relevant. These results are in agreement with the approximating solutions made in the previous sections for specific ranges of the model parameters.

\begin{figure}[bt]
    \centering
    \includegraphics[width=1\textwidth]{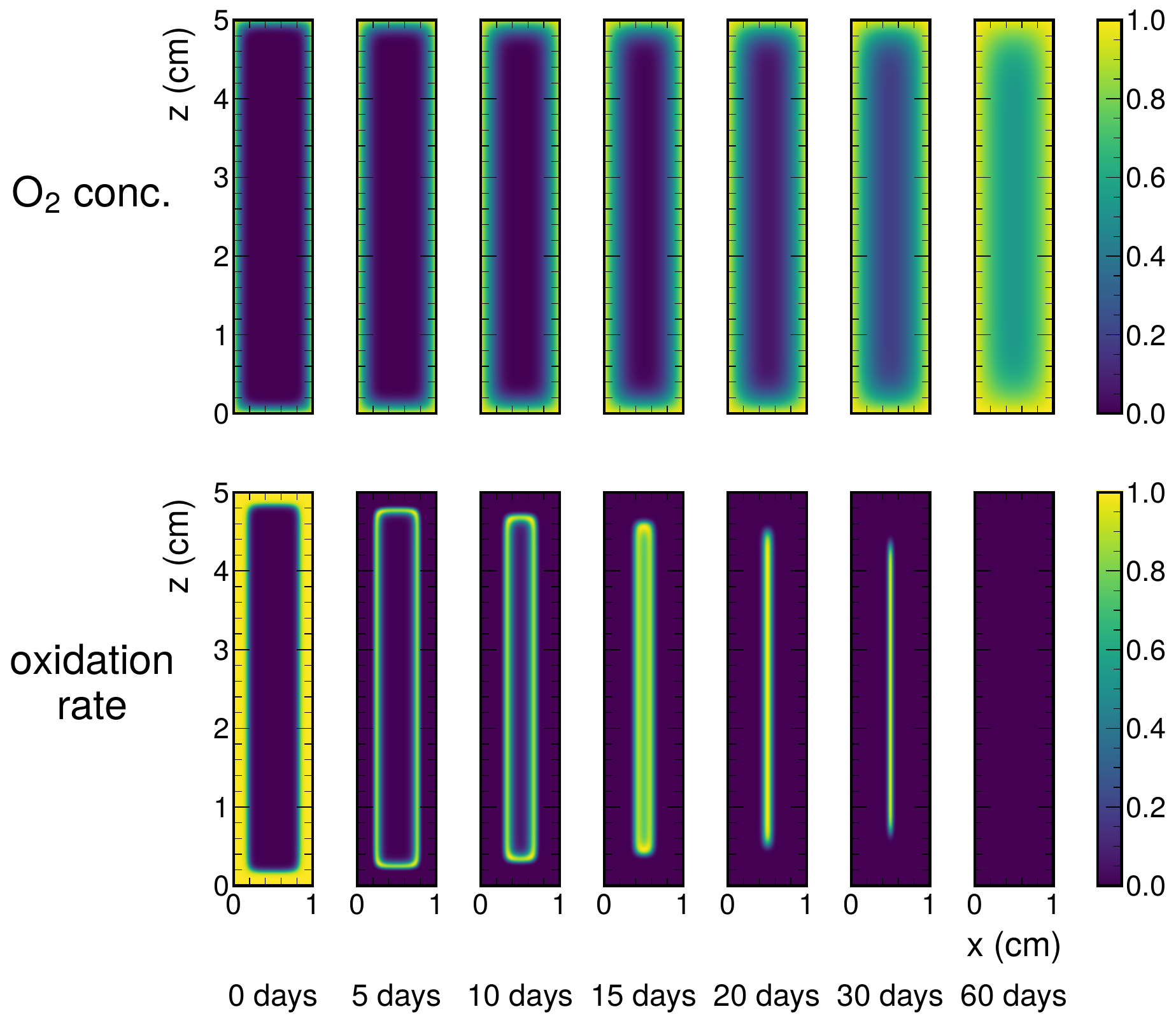}
    \caption{Distributions of the normalized \ch{O2} concentration and the oxidation rate for different points in time during the annealing process for the 3D finite differences simulation of the JAERI model. The input parameters are $k_1=5\cdot10^{-11}\,\mathrm{cm}^3\mathrm{s}^{-1}$, $k_2=2\cdot10^{-10}\,\mathrm{cm}^3\mathrm{s}^{-1}$, and $Y\doserate=0.6\,\mathrm{cm}^{-3}\mathrm{s}^{-1}$. At $t=0$, the irradiation stops and the annealing period begins.}
    \label{fig:sim_vis}
\end{figure}

\begin{figure}[bt]
    \centering
    \includegraphics[width=1\textwidth]{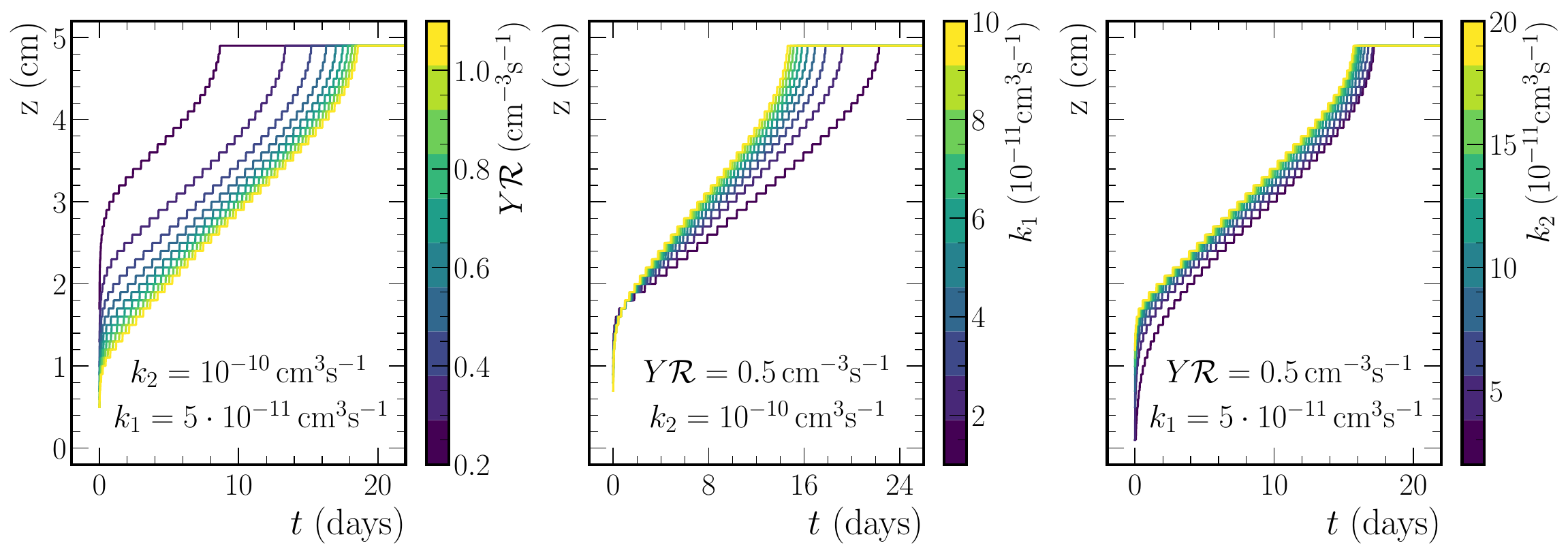}
    \caption{The 3D finite differences simulation of the JAERI model for various combinations of the input parameters $Y\doserate$, $k_1$, and $k_2$. In each plot, two of the parameters are kept constant and the third is varied over some range. The diffusion coefficient is $D=10^{-8}\,\mathrm{cm}^{2}\mathrm{s}^{-1}$ in all cases.}
    \label{fig:simulation}
\end{figure}

Understanding the dynamical evolution of the system requires a full simulation of Eqs.\ref{eq:radicals1} and \ref{eq:oxygen}. For that reason, a custom CUDA implementation of the model is used to create a finite differences simulation of the 3-dimensional sample. The simulation utilizes a $500\times100\times100$ grid for the $5\times1\times1\,$cm samples, and the input parameters are $D$, $Y\doserate$, $k_1$, and $k_2$. The diffusion coefficient $D$ is set to a value of $0.1\,\frac{\mathrm{(lattice\ units)^2}}{\mathrm{time\ steps}}$ in order to keep the simulation runtime manageable. This value is equivalent to $10^{-8}\,\mathrm{cm}^2\mathrm{s}^{-1}$ after considering that the experimental annealing time is approximately 30 days. The system is initialized with a nominal value of the oxygen concentration $C_0=1$ throughout the sample volume, and this value is kept steady at the outer boundary of the sample, thus imposing a Dirichlet boundary condition. Both the oxygen and the polymer radical concentrations are evolved for 10,000 time steps, simulating the irradiation. Subsequently, the $Y\doserate$ parameter is set to 0 and the system is evolved for another 10,000 steps, simulating the annealing period. Fig.~\ref{fig:sim_vis} 
shows the normalized oxygen concentration and oxidation rate for the slice at $y=50$ (middle of that dimension) over different times during the annealing process and for input parameters $k_1=5\cdot10^{-11}\,\mathrm{cm}^3\mathrm{s}^{-1}$, $k_2=2\cdot10^{-10}\,\mathrm{cm}^3\mathrm{s}^{-1}$, and $Y\doserate=0.6\,\mathrm{cm}^{-3}\mathrm{s}^{-1}$. The first time step, at $t=0$, represents the time when the irradiation is turned off. The snapshots of the system at that point in time reveal that the rate of oxidation is very high in the periphery of the sample and drops suddenly to 0 at a certain depth. At the same time, the \ch{O2} concentration is very low throughout the sample and practically 0 in the area with no oxidation rate. As the annealing progresses, the oxygen diffuses inside the sample, and the \ch{O2} concentration increases gradually. The oxidation rate is high only in a short band inside the sample, which corresponds to the area that has sufficient oxygen and radicals to sustain the reaction. This band moves towards the center of the sample with increasing time, until the time when all the radicals have been oxidized and the sample is considered fully annealed. Fig.~\ref{fig:simulation} 
shows a comparison of the annealing process for a variety of combinations of the parameter values. The annealing is represented by the depth $z$ of the area with a high oxidation rate at the middle of the sample $(y=50, z=250)$. For each one of the three plots, two out of three parameters are kept stable and the third parameter is varied over a certain range. The dose rate times the specific rate of radical formation, expressed as the combined parameter $Y\doserate$, affects significantly the initial depth, the duration, and the shape of the annealing curve. Lower values tend to place the initial position deeper into the sample and make the annealing curve look less linear. The rate of radical crosslinking $k_1$ mainly affects the annealing duration with lower values extending its length. Finally, the rate of radical oxidation $k_2$ has a significant effect on the initial position, which also affects the annealing duration.

\subsection{\label{subsec:alt}Photo-oxidation model}

A different approach is followed by Cunliffe and Davis in \cite{cunliffe} for describing the dynamics between oxygen diffusion and light-induced oxidation (which are applicable to ionizing radiation as well). This description is based on a similar chemical description of the phenomenon that is accounted for by the addition of a reaction rate term $R(C(x,t))$ to the diffusion equation (Eq.~\ref{eq:ficks2nd})
\begin{equation}\label{eq:oxRed}
    \frac{\partial C(x,t)}{\partial t} = D\frac{\partial^2 C(x,t)}{\partial x^2} - R(C(x,t)),
\end{equation}
where $R(C(x,t))$ is the rate of oxygen consumption. For a free radical reaction chain with initiation, propagation, and termination steps \cite{FURNEAUX1981431,10.1007/BFb0051109,TF9484400669,QR9490300001,QR9540800147}, this rate can be described by the formula
\begin{eqnarray}\label{eq:consRate}
    R(C(x,t))&=&\frac{c_1'C(x,t)}{c_2'C(x,t)+c_3'}\nonumber\\
    &=&\frac{c_1C(x,t)}{c_2C(x,t)+1},
\end{eqnarray}
where $c_1'$, $c_2'$, $c_3'$, and therefore $c_1\equiv\frac{c_1'}{c_3'}$ and $c_2\equiv\frac{c_2'}{c_3'}$, can be expressed in terms of the rate constants for the initiation, propagation, and termination steps. By comparing Eq.~\ref{eq:consRate} with Eq.~\ref{eq:oxygen}, it is evident that $-k_2[\ch{P^.}]C(x,t)$ is replaced by $R(C(x,t))$, and the creation and termination of the radicals are modeled via the propagation and termination. Therefore, this model features more simplified handling of the radical processes. The steady-state differential equation becomes
\begin{equation}\label{eq:diffEqAlt}
    D\frac{\partial^2 C(x)}{\partial x^2} = \frac{c_1C(x)}{c_2C(x)+1},
\end{equation}
and after reusing $y\equiv\frac{C}{C_0}$, $\lambda\equiv\frac{x}{l}$,
\begin{eqnarray}\label{eq:diffEqAltFinal}
    \frac{d^2y}{d\lambda^2}&=&\frac{\left(\frac{c_1l^2}{D}\right)y}{c_2C_0y+1}\nonumber\\
    &=&\frac{Ey}{Fy+1},
\end{eqnarray}
where $E\equiv\frac{c_1l^2}{D}$, and $F\equiv c_2C_0$. The boundary conditions remain the same as before.

\subsubsection{\label{subsubsec:alt_cond1}If $F\ll1$}
For $F$ to have a very small value, either the initial concentration of oxygen $C_0$ needs to be very low or the rate constants for the radical oxidation process and the termination process for oxidized radicals need to be much lower than the rate constant for the termination of non-oxidized radicals. In this case, the reaction rate is proportional to $C$. For small $E$ values, which correspond to a combination of small sample thickness, low radical formation rates, low dose rates, and low radical oxidation rates, the oxygen diffusion offsets the reaction sufficiently and the decrease of $C$ towards the center of the sample is small. For larger values of $E$, the decrease of $C$ becomes more pronounced.

\subsubsection{\label{subsubsec:alt_cond2}If $Fy\gg1$ for all $y$}
In this case, there is sufficient oxygen concentration throughout the sample and the rate constants for the processes that involve oxygen are sufficiently high therefore the reaction rate is approximately constant throughout the sample. For larger values of $E$, the demand for oxidation increases and the oxygen concentration can be depleted within an inner region of the sample.

\subsubsection{\label{subsubsec:alt_cond3}If $Fy\approx1$ for some $y$}
Considering the case where $F\gg1$, the reaction rate should be approximately constant as long as $Fy\gg1$ holds. However, the validity of this statement depends as well on the value of $E$. Specifically, for small values of $E$, the condition $Fy\gg1$ is valid for all values of $y$. As the value of $E$ increases, oxygen concentration falls more and more rapidly with depth into the sample. For large values of $E$, a balance is created between the oxygen concentration and the oxidative part of the reaction chain in some regions of the sample, and as a result, there are some $y$ for which $Fy\approx1$. In this case, $C$ has a plateau region with almost constant oxygen concentration, and then at some critical depth $\lambda_c$ the concentration collapses dramatically to very low values. The effect is more pronounced for larger values of $E$ and $F$ provided that $E$ is sufficiently larger than $F$. The critical depth can be estimated by beginning with condition $Fy\gg1$. Then
\begin{equation}
    \frac{d^2y}{d\lambda^2}=\frac{E}{F}.
\end{equation}
After integrating twice and applying the boundary conditions,
\begin{equation}\label{eq:largeABsolution}
    y=1-\frac{E}{8F}+\frac{E\lambda^2}{2F}.
\end{equation}
Considering the critical condition when $y$ falls below $y_c$ given by $Fy_c\approx1\Leftrightarrow y_c\approx\frac{1}{F}$ in the middle of the sample,
\begin{equation}
    \frac{E}{8F}\geq1-\frac{1}{F}.
\end{equation}
For large $F$ values,
\begin{equation}
    \frac{E}{8F}=\frac{c_1l^2}{8Dc_2C_0}\geq1.
\end{equation}
This critical value $y_c$ is very small and therefore the roots of Eq.~\ref{eq:largeABsolution} can give an approximation of the location of the sharp decreases:
\begin{equation}
    \lambda_c^2 = \frac{1}{4} - \frac{2F}{E} = \frac{1}{4} - \frac{2c_2DC_0}{c_1l^2}.
\end{equation}
Writing for $z_c$, which is measured from the surface of the sample
\begin{equation}
    z_c^2 = \frac{2c_2DC_0}{c_1}
\end{equation}
This formula is very similar to Eq.~\ref{eq:depth} provided that the ratio $c_1/c_2$ is approximately equal to $Y\doserate$.

\subsubsection{\label{subsubsec:altnNumSol}Numerical solution}

\begin{figure}
    \centering
    \includegraphics[width=1\textwidth]{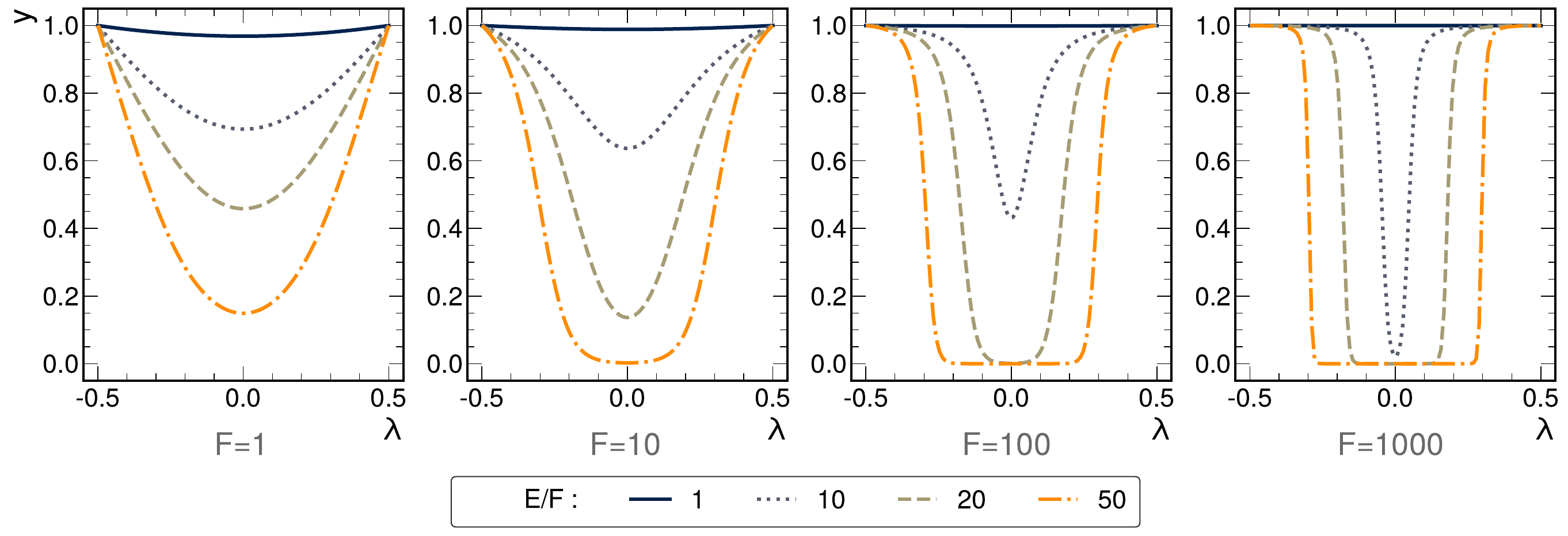}
    \caption{\label{fig:wide}Numerical solutions to Eq.~\ref{eq:diffEqAltFinal} for various combinations of values for the parameters $E\equiv\frac{c_1l^2}{D}$ and $F\equiv c_2C_0$. The parameter $E$ is expressed through its ratio over $F$. The normalized reaction rate $R/R_0$ is shown for the entire normalized length $\lambda$. For the code used see Ref.~\cite{repository}.}
\end{figure}

Equation~\ref{eq:diffEqAltFinal} can be solved numerically using \verb|scipy|'s ODE solver for boundary value problems. The full code can be found in \cite{repository}. Fig.~\ref{fig:wide} shows solutions of the normalized reaction rate $R/R_0$ for different values of the model parameters $A$ and $B$. Parameter $B$ controls the shape of the solution, which forms shoulder regions in the outer parts of the sample and drops steeply in the middle for values $\gtrsim10$. The ratio $A/B$ defines the depth where the transition between the two regions occurs.

\begin{figure}[bt]
    \centering
    \includegraphics[width=1\textwidth]{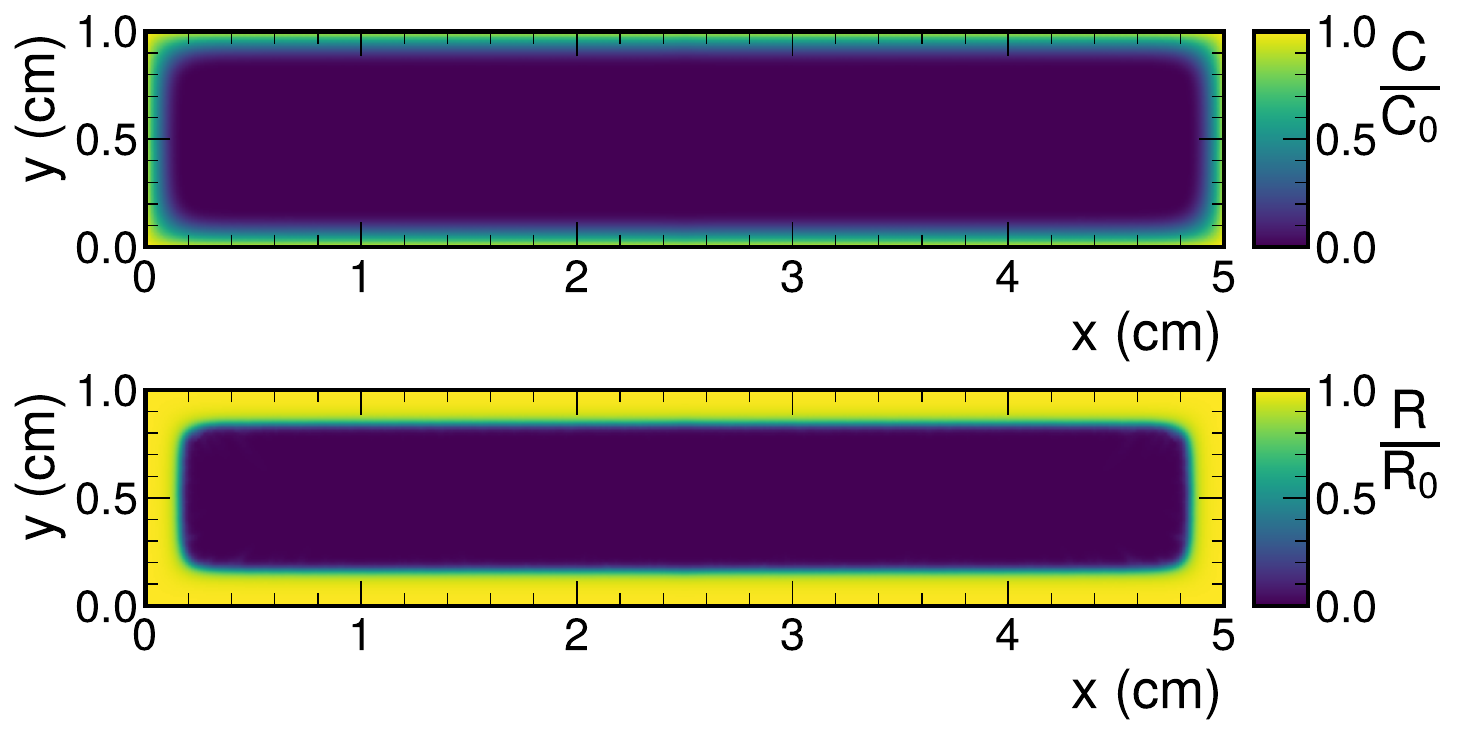}
    \caption{The solution to the 2D PDE for the normalized concentration and reaction rate at steady state for a scintillator that is $1\,\textrm{cm}\times1\,\textrm{cm}\times5\,\textrm{cm}$. For the code used see Ref.~\cite{repository}.}
    \label{fig:2Dpde}
\end{figure}

For the sake of completeness, Eq.~\ref{eq:diffEqAlt} was also solved for the two-dimensional case that matches the physical size of our samples, which are described in detail in Section~\ref{sec:exper}. The dynamic system was evolved over time on a $350\times70$ grid until a steady solution was reached. In Fig.~\ref{fig:2Dpde},
the solution is shown for the normalized concentration $C/C_0$ and the normalized reaction rate $R/R_0$ for a choice of the model parameters $c_1=c_2=100$ and $D=0.01$. These values were chosen because they produce a steady state in which the oxygen concentration drops significantly very close to the surface of the material, and the reaction rate is high close to the surface but then suddenly drops at some critical depth $z_c$.

\section{\label{sec:refrIndexTheory}Refractive index theory}
The refractive index $n$ for a material can be expressed as
\begin{equation}
    n=\sqrt{\epsilon_r\mu_r},
\end{equation}
where $\epsilon_r$ and $\mu_r$ are the relative permittivity and permeability, respectively. For non-ferromagnetic materials, the relative magnetic permeability is very close to unity. So for plastic scintillators, the formula
\begin{equation}
    n\simeq\sqrt{\epsilon_r}
\end{equation}
is a good approximation.

A connection between $\epsilon_r$ and the microscopic description of a material can be established by the Clausius-Mossoti equation \cite{doi:10.1021/j150334a007}
\begin{equation}
    \frac{\epsilon_r-1}{\epsilon_r+2}=\frac{N\alpha}{3\epsilon_0},
\end{equation}
where $N$ is the number density of molecules and $\alpha\equiv\frac{\norm{\mathbf{p}}}{\norm{\mathbf{E}}}$ is the molecular polarizability of the material, which is defined as the tendency of the molecules to acquire a dipole moment $\mathbf{p}$ when exposed to an external electric field $\mathbf{E}$. 

This can be rewritten for the refractive index $n$ in a form known as the Lorentz-Lorenz equation \cite{Kragh_2018,doi:10.1002/andp.18802470905,Lorentz1936}
\begin{equation}
    \frac{n^2-1}{n^2+2}=\frac{4\pi}{3}N\alpha_m,
\end{equation}
where $\alpha_m$ is the average molecular polarizability volume (that is the polarizability expressed in cgs in order to give the convenient dimension of volume). Therefore, the refractive index can be predicted using only one microscopic quantity, $\alpha_m$.

The average molecular polarizability depends on the atomic polarizabilities of the atoms that form the substance's molecules as well as on the bond strengths and the molecular structure. This fact was realized by chemists in the late 18th century trying to associate the structure of organic molecules with their refractive indices. Br\"uhl pioneered a method \cite{bruhl1887influence, doi:10.1002/cber.18910240232} to predict the molecular refractivity
\begin{equation}\label{molRefr}
    MR\equiv\frac{n^2-1}{(n^2+2)N}
\end{equation}
by assigning a weighted sum of the atomic refractivities to it. He calculated the atomic refractivities for the atoms that are most commonly found in organic molecules and the additional factors associated with double and triple bonds, and he was able to predict the correct structure for benzene. Table~\ref{tab:table1}
\begin{table}[bt]
\centering
\caption{\label{tab:table1}
Br\"uhl's weights. The sum of these weights gives the molecular refractivity for a compound.
}
\begin{tabular}{lcccccc}
\toprule
\textrm{Atom/bond} & \ch{C} & \ch{H} & \ch{O} & \ch{C=C} & \ch{C+C} & \ch{C=O}\\
\midrule
\textrm{weight} $r_i$ & 2.48 & 1.04 & 1.58 & 1.78 & 2.18 & 0.76\\
\bottomrule
\end{tabular}
\end{table}
shows the values that Br\"uhl derived for the weights of the constituents of organic compounds. Calculations according to Br\"uhl theory are shown in Table~\ref{tab:table2}
\begin{table}[bt]
\centering
\caption{\label{tab:table2}
Br\"uhl theory predictions and experimental values for three common organic substrate materials.
}
\begin{tabular}{lccccc}
\toprule
\textrm{Polymer} & $\sum r_i$ & $P$ (g/mol) & $d$ (g/mL) & $n_{\textrm{Br\"uhl}}$ & $n_{\textrm{experiment}}$\\
\midrule
\textrm{PMMA} & 26.22 & 100.12 & 1.15 & 1.510 & 1.49\\
\textrm{PVT} & 38.06 & 118 & 1.023 & 1.574 & 1.58\\
\textrm{PS} & 33.50 & 104.15 & 1.05 & 1.591 & 1.59\\
\bottomrule
\end{tabular}
\end{table}
, where the values for the molecular weight $P$ of the monomers, the density $d$, and the experimental value of the refractive index $n_{\textrm{experiment}}$ are retrieved from \cite{polymerdatabase, eljendata}. Subsequently, the same equation can predict the difference $\delta n$ in the refractive index when the chemical composition of the material changes as is expected to happen during irradiation. For example, substituting two hydrogen atoms for a double carbon bond in styrene is expected to change the refractive index by $\delta n = 0.008$ while replacing two hydrogen atoms in the same monomer with a carboxyl bond gives $\delta n = -0.079$. Therefore, it is expected that the presence or absence of oxygen during the irradiation is going to affect the refractive index since the neutralization of the radicals will differ.

The characterization of materials based on their refractive index is often performed by fitting empirical formulas, that describe the dispersion with light wavelength, to experimental data. The most popular formulas are the Cauchy equation
\begin{equation}\label{eq:cauchy_eq}
    n(\lambda) = A + \frac{B}{\lambda^2} + \frac{C}{\lambda^4} + ... \approx A + \frac{B}{\lambda^2},
\end{equation}
where $A$ and $B$ are material-dependent coefficients whose values are determined experimentally, and the Sellmeier equation \cite{sellmeier}
\begin{equation}\label{eq:sellmeier_eq}
    n^2(\lambda) = 1 + \sum_{i=1}^{N}\frac{B_i\lambda^2}{\lambda^2-C_i},
\end{equation}
where $B_i$ and $C_i$ are the Sellmeier coefficients and they are experimentally determined. This equation is considered an improvement over Cauchy's formula as it is better at modeling the refractive index in a broad wavelength range from the infrared to the ultraviolet part of the spectrum. Usually up to third-order coefficients are used to characterize glasses, although keeping two or even one order is sufficient for many applications.

\section{\label{sec:exper}Experimental setup and procedures}

\subsection{\label{subsec:irrSec}Samples and irradiations}
All scintillators discussed in this work are rectangular rods of dimension $1\,\textrm{cm}\times1\,\textrm{cm}\times5\,\textrm{cm}$, supplied by Eljen Technology. The substrate material is either polystyrene (PS) or polyvinyltoluene (PVT), while the primary and secondary fluors are p-Terphenyl and POPOP-type, respectively. These scintillators are commercially available from Eljen Technology as EJ-200 with PVT substrate material. The PS-based scintillators were prepared by Eljen specifically for these studies as a custom request. 

The rods were irradiated with different dose rates at room temperature in three irradiation facilities. Irradiations from $80.6\,$Gy/h to $3.9\,$kGy/h with integrated doses of $70\,$kGy were performed at the National Institute for Standards and Technology (NIST) in Gaithersburg, MD, using $\ch{^{60}Co}$. Lower dose rate irradiations from 3.0 to $9.8\,$Gy/h with total doses ranging from 12.6 to $42\,$kGy were performed at Goddard Space Flight Center, using also $\ch{^{60}Co}$. The lowest dose rate irradiations were performed at the GIF++ facility at CERN \cite{PFEIFFER201791} using $\ch{^{137}Cs}$ with a dose rate of $2.2\,$Gy/h to a total dose of $13.2\,$kGy. The dose and dose rate uncertainties are 1.2\% for the NIST irradiations and 10\% for the Goddard and GIF++ irradiations. All samples were stored in a dark container after irradiation until annealing was complete.

\subsection{\label{subsec:postIrr}After irradiation}

\begin{figure}[hbtp]
    \centering
    \includegraphics[width=0.49\textwidth]{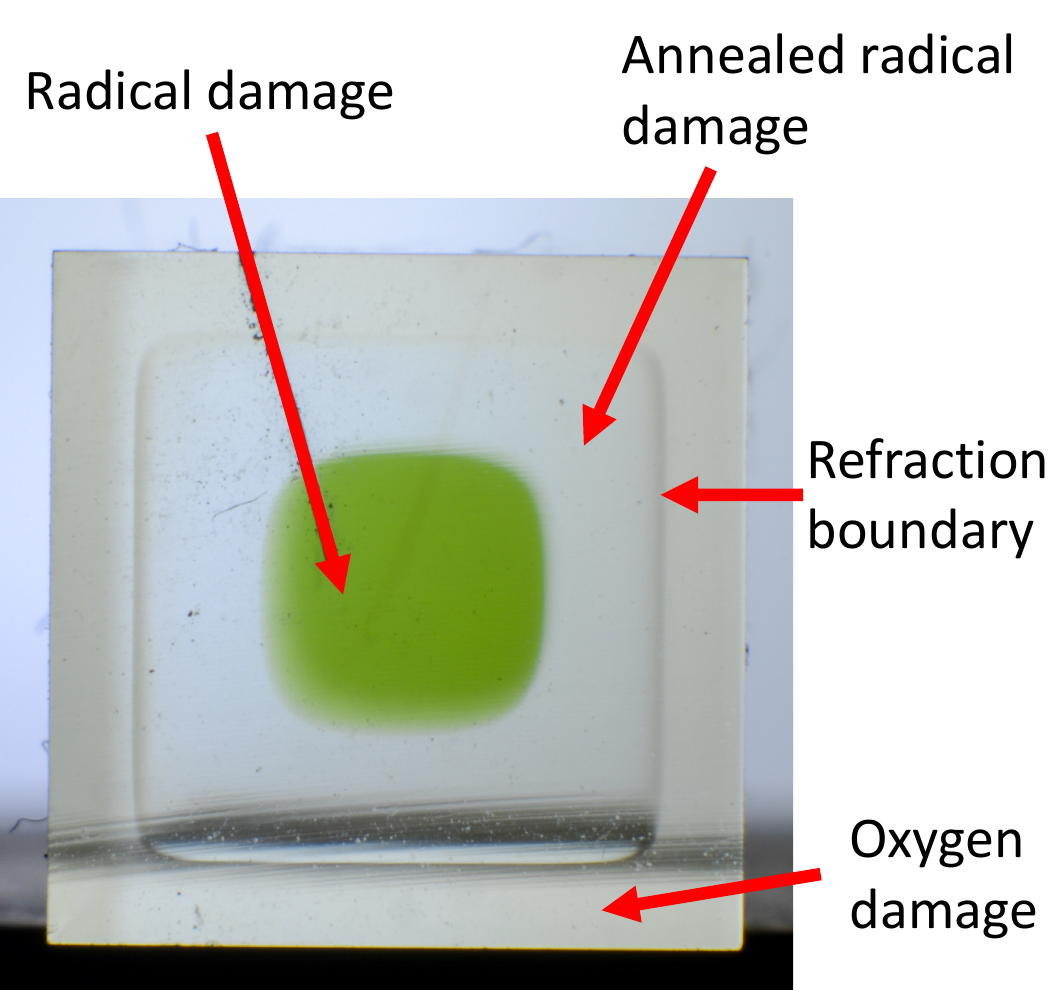}
    \includegraphics[width=0.49\textwidth]{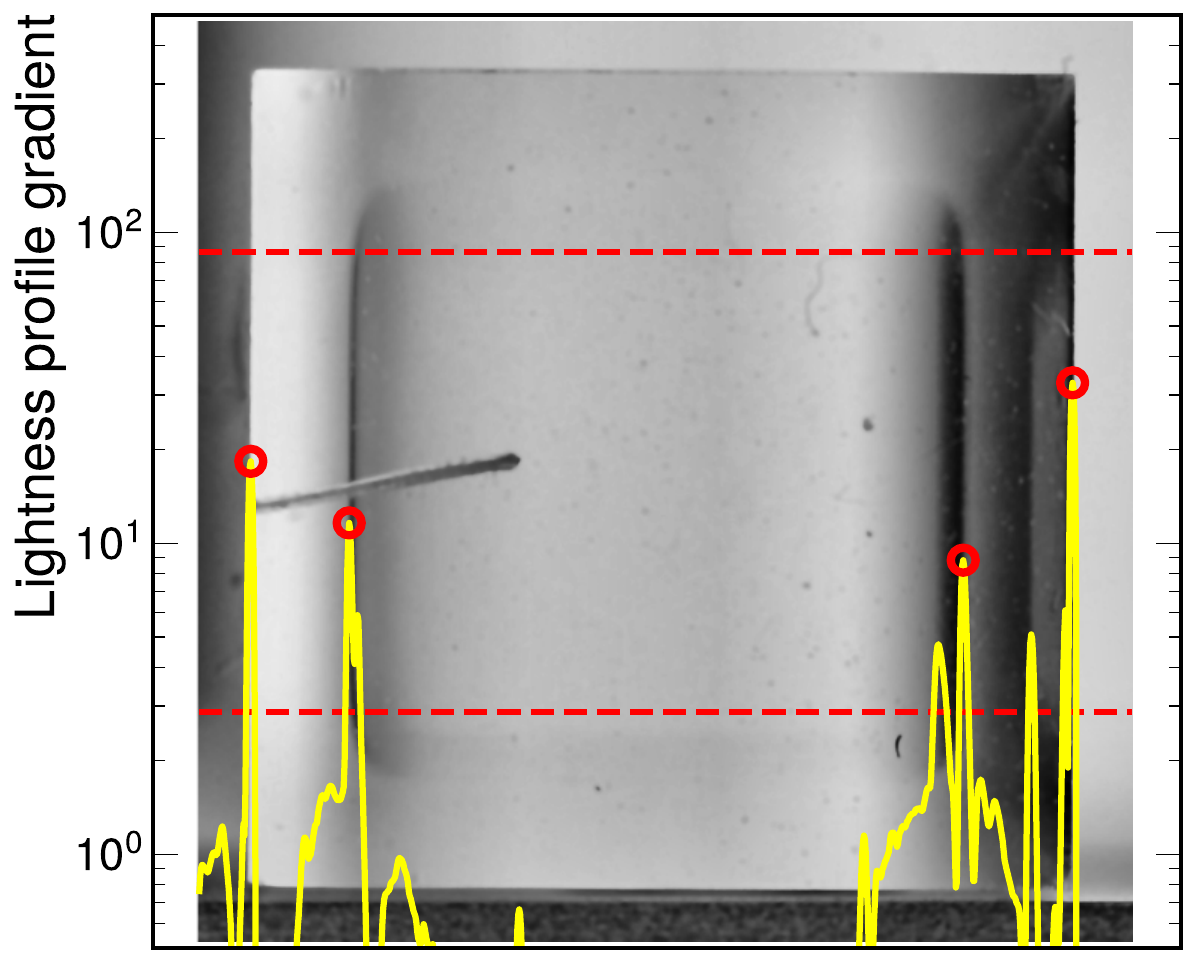}
    \caption{[Left] One of the irradiated scintillator samples with a visible refractive index change boundary 15 days after the irradiation end. Irradiation was performed using $\ch{^{60}Co}$ at NIST to a total dose of $70\,$kGy at a dose rate of $460\,$Gy/h. [Right] The processed lightness channel of a scintillator sample. The gradient of the profile is shown in yellow color, the range used for the profile is defined by red dashed horizontal lines, and the peaks are identified with red circles.}
    \label{fig:sample}
\end{figure}

Irradiated plastic scintillator acquires distinctive optical features that can be used to indicate the recovery status or even measure any abrupt changes in the reaction rate between oxygen and the radiation-induced radicals. Visual inspection of the samples immediately after exiting the irradiator reveals an internal region with a characteristic green coloration from radical color centers that is surrounded by a boundary, which indicates a rapid change in the refractive index of the material. During the recovery process, the green coloration gradually shrinks toward the center of the sample until it eventually disappears. The time needed for the completion of this process varies by the substrate material and depends on the storage conditions. For the samples discussed in this work, the average time to full recovery is between 1 and 2 months. This process is already well documented in literature and is known as scintillator "bleaching" or annealing. Provided enough oxygen is available during recovery, all radicals will get oxidized and therefore the green coloration will get neutralized. After complete annealing, the only visual remnant of the irradiation is the boundary at which the refractive index of the material changes sharply. Fig.~\ref{fig:sample}
[Left] shows the $1\times1\,$cm face of a PS scintillating rod 15 days after the end of its irradiation with $460\,$Gy/h for an integrated dose of $70\,$kGy. 

\subsection{\label{depthMeas}Refractive index boundary depth}
The refractive index change is present in all scintillator rods irradiated with dose rates $>10\,$Gy/h. The refractive index boundary depths were measured using an OpenCV-based method \cite{opencv_library} with photographs and video frames of the samples taken in front of high-contrast backgrounds. The method consists of the following steps: 
\begin{enumerate}
    \item Select pictures/frames with the cleanest views of the boundaries.
    \item Process with a bilateral filter to reduce noise and dust.
    \item Extract the lightness channel.
    \item Improve contrast with Contrast Limited Adaptive Histogram Equalization (CLAHE) \cite{Ketcham1974ImageET, HUMMEL1977184}.
    \item Extract the lightness profile along the axis used for depth calculation.
    \item Calculate the gradient of the lightness profile to reveal any abrupt changes.
    \item Apply a Savitzky–Golay filter \cite{doi:10.1021/ac60214a047} for noise removal.
    \item Find the positions and widths of the gradient peaks.
    \item Calculate the image scale in mm/px and the depth of the refractive index boundaries using the peak positions.
    \item Use the peak widths and the variance over a few different pictures to estimate the uncertainties. 
\end{enumerate}
Fig.~\ref{fig:sample} [Right] shows the lightness profile gradient overlaid over the processed lightness channel for one of the samples. The discovered peaks that are later used for calculating the image scale and the boundary depth are shown with red circles. The code and the data that can fully reproduce all the results can be found in Ref.~\cite{repository}.

\subsection{\label{indexMeas}Refractive index measurement}

\begin{figure}[hbtp]
    \centering
    \includegraphics[width=1.\textwidth]{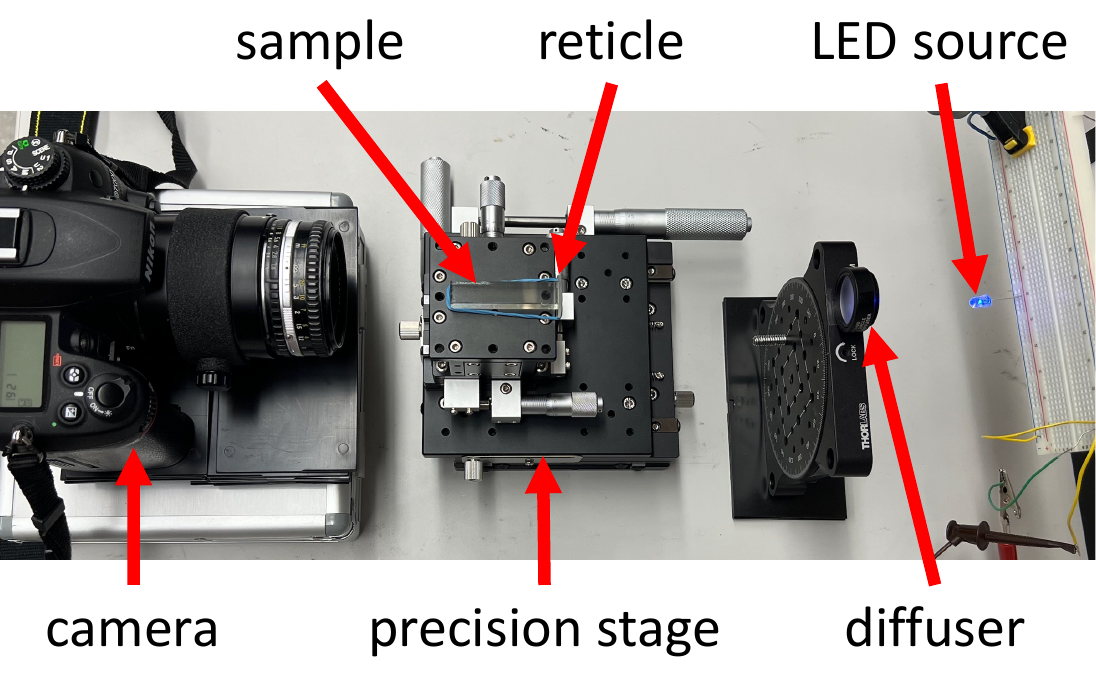}
    \caption{The setup for measurements of the index of refraction using a Nikon D7000 digital camera, an MPositioning precision stage with a sample adjacent to a Boli Optics $10\times10$\,mm/$100\times100$ reticle, a Thorlabs $20\degree$ circle pattern diffuser, and a $470\,$nm LED source.}
    \label{fig:indexMeasSetup}
\end{figure}

The refractive index of the scintillator rods was measured using the setup shown in Fig.~\ref{fig:indexMeasSetup}. 
A Nikon D7000 DSLR was aligned with the axis of an MPositioning precision translation stage. The rods were placed on the translation stage with a Boli Optics $0.1\,$mm division net grid reticle attached to the face opposite the camera. The rod-reticle system was illuminated evenly from behind using an LED of known wavelength with a Thorlabs $20\degree$ circle pattern engineered diffuser. 

\begin{figure}
    \centering
    \includegraphics[width=1.\textwidth]{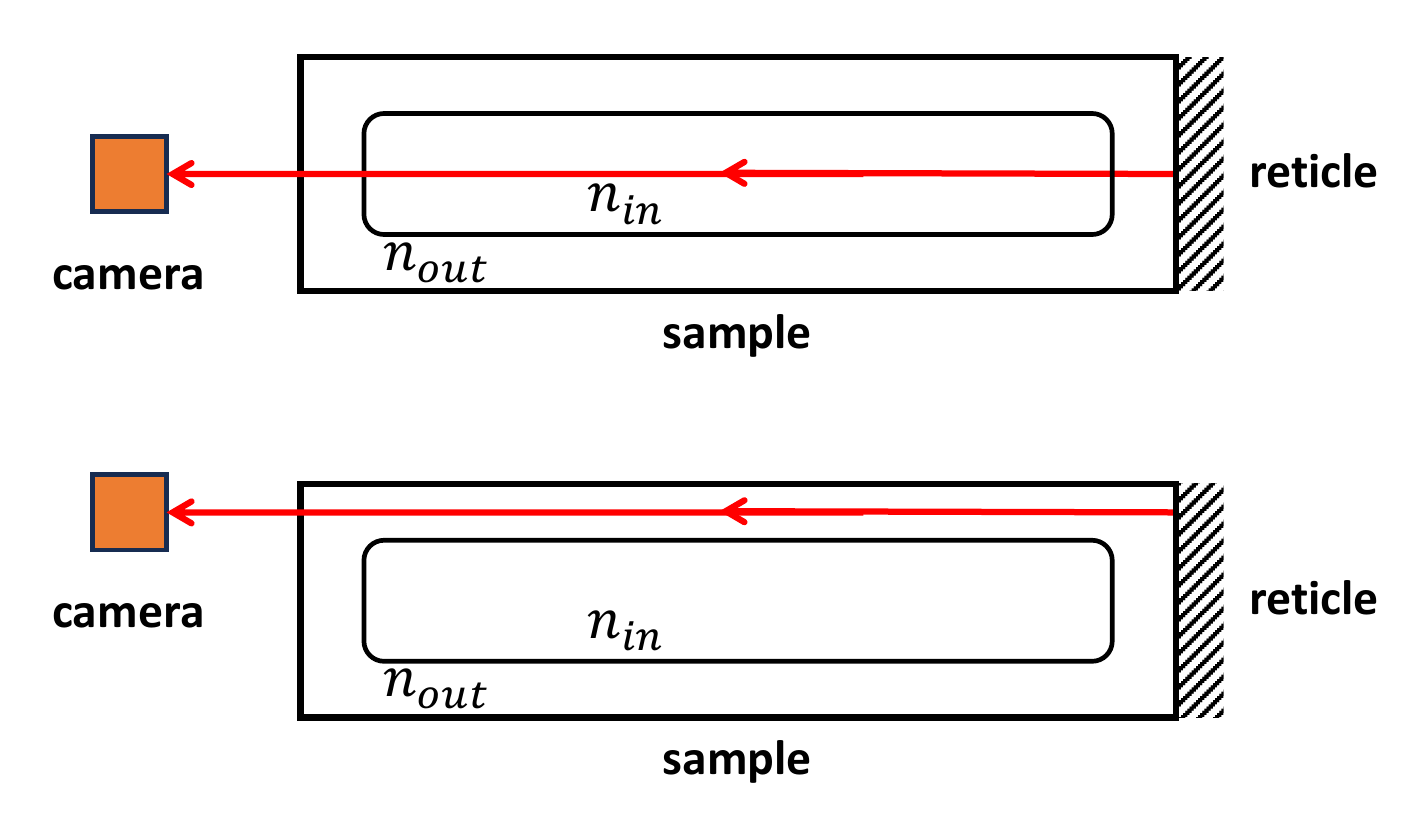}
    \caption{A schematic showing the configurations for measuring the refractive indices in samples with two regions. The index of the inner region $n_{in}$ is measured using the configuration shown at the top in which the light passing through the middle of the sample is used. The index of the outer region $n_{out}$ is measured according to the configuration shown at the bottom, which uses light passing through the periphery of the sample thus avoiding the inner region. The fact that light traversing the inner region also traverses some of the material in the outer region requires a correction term to the inner region's measured index, as described in the text.}
    \label{fig:indexRegionsSchematic}
\end{figure}

The camera lens focus was fixed on the front face of the rod. Then, the precision stage was translated forward up to the distance at which the divisions of the reticle at the other end of the rod appeared to be in focus. The total distance traversed from the front to the apparent end of the rod was the measured apparent depth $L'$. This procedure was performed for both the inside and outside regions of the rods as is shown in Fig.~\ref{fig:indexRegionsSchematic}. 
The rod lengths $L$ were measured using a caliper. For a sample without an internal boundary, the index of refraction is given by~\cite{Jaladri_2020}
\begin{equation}\label{eq:refr_index_1}
    n_{in} = \frac{L}{L'}.
\end{equation}
Accounting for two regions with different refractive indices $n_{in}$ and $n_{out}$ inside the rod, the index of the internal region is
\begin{equation}\label{eq:refr_index_2}
    n_{in} = \frac{(L-2d)n_{out}}{L'n_{out}-2d},
\end{equation}
where $d$ is the depth of the boundary, and $n_{out}$ is measured using Eq.~\ref{eq:refr_index_1} in the part of the sample where only the outer region is visible.

\begin{figure}[hbtp]
    \centering
    \includegraphics[width=0.49\textwidth]{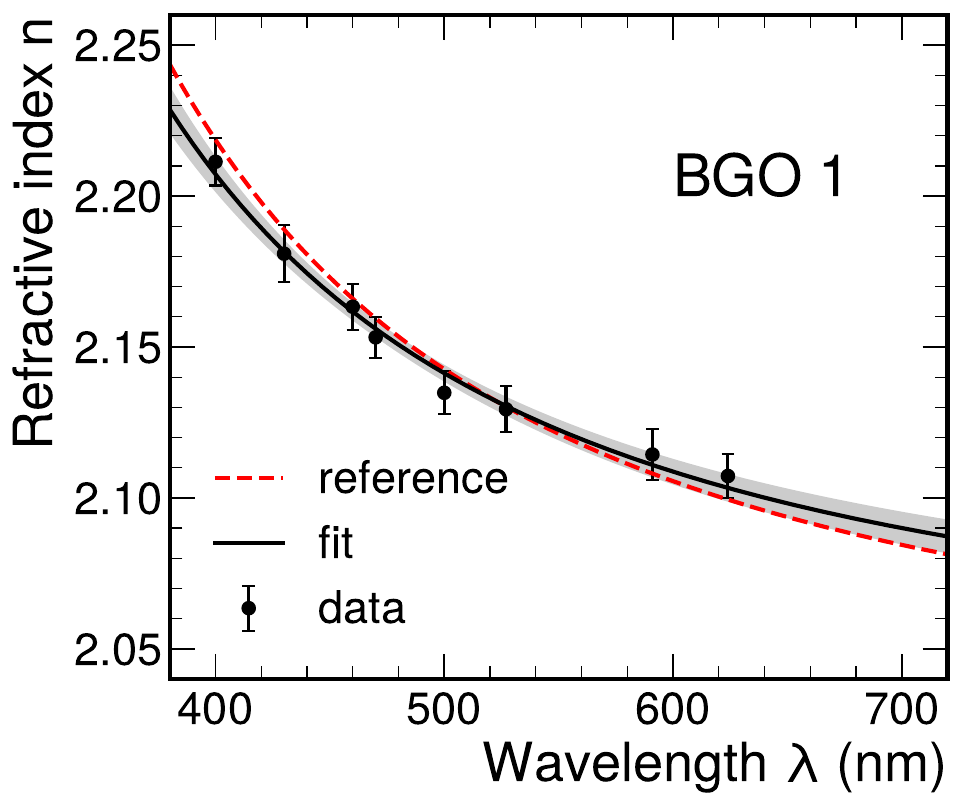}
    \includegraphics[width=0.49\textwidth]{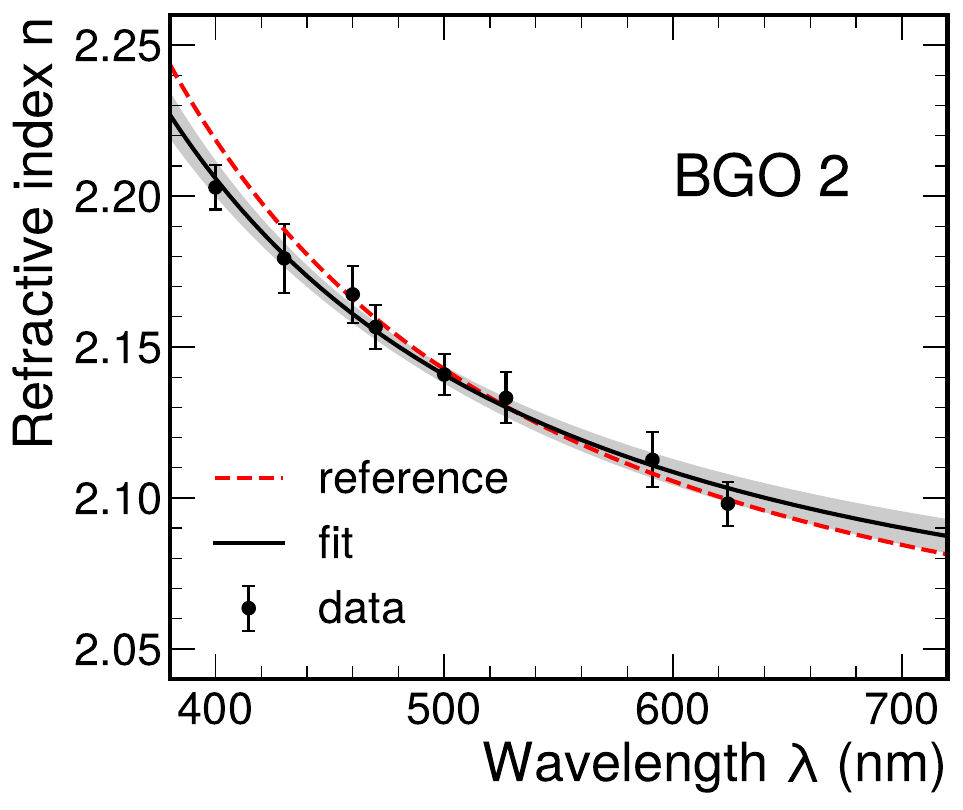}
    \caption{A validation of the refractive index measurement method performed against the Sellmeier equation for two BGO crystal samples. The plots show measurements of the index of refraction versus light wavelength. The red curve is the Sellmeier equation using the first-order Sellmeier coefficients that can be found in the literature. A fit of the data to the first-order Sellmeier equation is shown in black. The $1\sigma$ band is shown in gray color around the fit line.}
    \label{fig:sellmeier}
\end{figure}

The method was validated using two \ch{Bi4Ge3O12} (BGO) crystal samples. The validation was performed by measuring the index of the material at eight different wavelengths and then comparing their distribution with the relationship predicted by Eq.~\ref{eq:sellmeier_eq}. The reference values for $B_1$ and $C_1$ for BGO crystals can be found in Ref.~\cite{rii}. Fig.~\ref{fig:sellmeier} 
shows the comparison between the reference Sellmeier equation and a fit of the same equation to our data for both samples. The fitted lines match reasonably well with the predictions throughout the visible light spectrum.

\section{\label{sec:results}Results}
\begin{figure}[hbtp]
    \centering
    \includegraphics[width=1\textwidth]{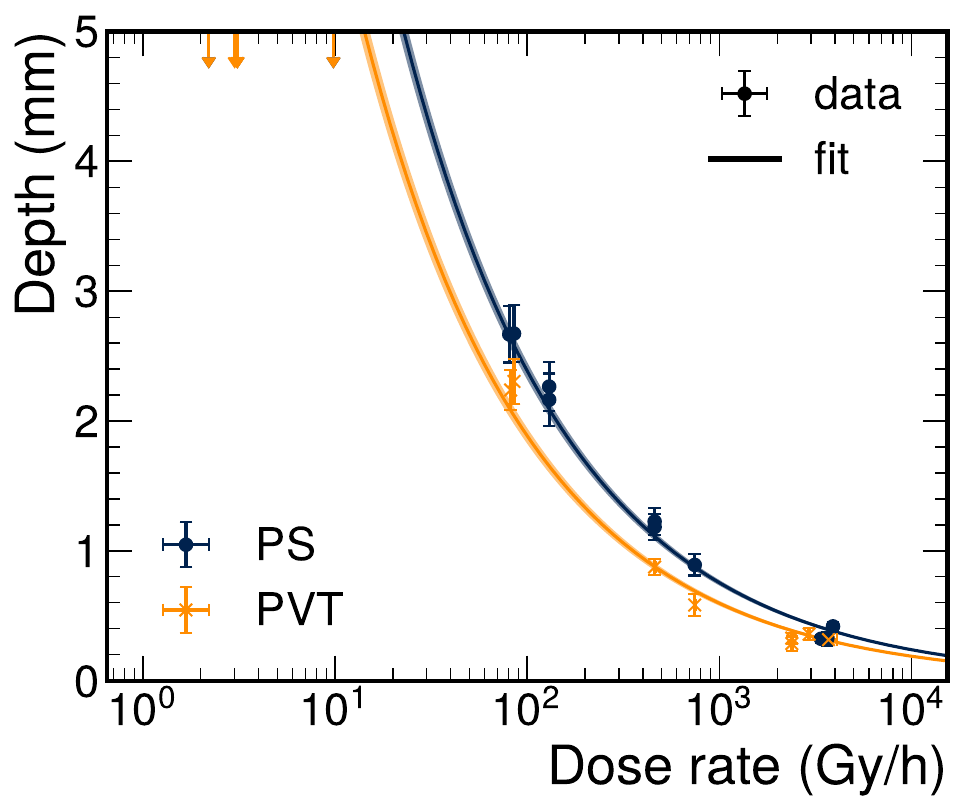}
    \caption{The measured depths of the refractive index change boundaries for irradiated PS and PVT scintillator samples as a function of dose rate. Irradiations at $2.2\,$Gy/h were done using a $\ch{^{137}Cs}$ source, the rest using $\ch{^{60}Co}$. The samples that were irradiated with dose rates lower than $80\,$Gy/h do not have a visible boundary indicating the depth is larger than half the sample thickness. For that reason, these samples are represented by lower pointing arrows at $5\,$mm whose length indicates the uncertainty of having a boundary so close to the surface that is impossible to measure or observe. The data above $80\,$Gy/h are fitted with a function of the form $z=\frac{A}{\sqrt{R}}$. The $1\sigma$ bands are also shown around the fit lines.}
    \label{fig:depth}
\end{figure}

Fig.~\ref{fig:depth} 
shows all measurements of depth versus the dose rate for PS and PVT rods. As expected by the hypothesis that this boundary is related to the oxygen penetration depth, the depth reduces with increasing dose rate because the oxygen diffusion is not enough to replenish the oxygen molecules consumed by the radicals. The data points are fitted with an equation of the form $z=A/\sqrt{\doserate}$ since this is the expected formula according to Eq.~\ref{eq:depth}. The values of the parameter $A$ are estimated to be $A_{\mathrm{PS}}=23.8\pm0.7\,\mathrm{mm}\cdot\mathrm{h}^{1/2}\mathrm{Gy}^{-1/2}$ for PS and $A_{\mathrm{PVT}}=18.8\pm0.6\,\mathrm{mm}\cdot\mathrm{h}^{1/2}\mathrm{Gy}^{-1/2}$ for PVT. Converting these values to dose rates that allow full oxygen penetration to $5\,$mm (half the sample width) gives $R^{full}_{\mathrm{PS}}=22.7\pm1.3\,$Gy/h for PS and $R^{full}_{\mathrm{PVT}}=14.2\pm0.9\,$Gy/h for PVT.

\begin{figure}
    \centering
    \includegraphics[width=1.0\textwidth]{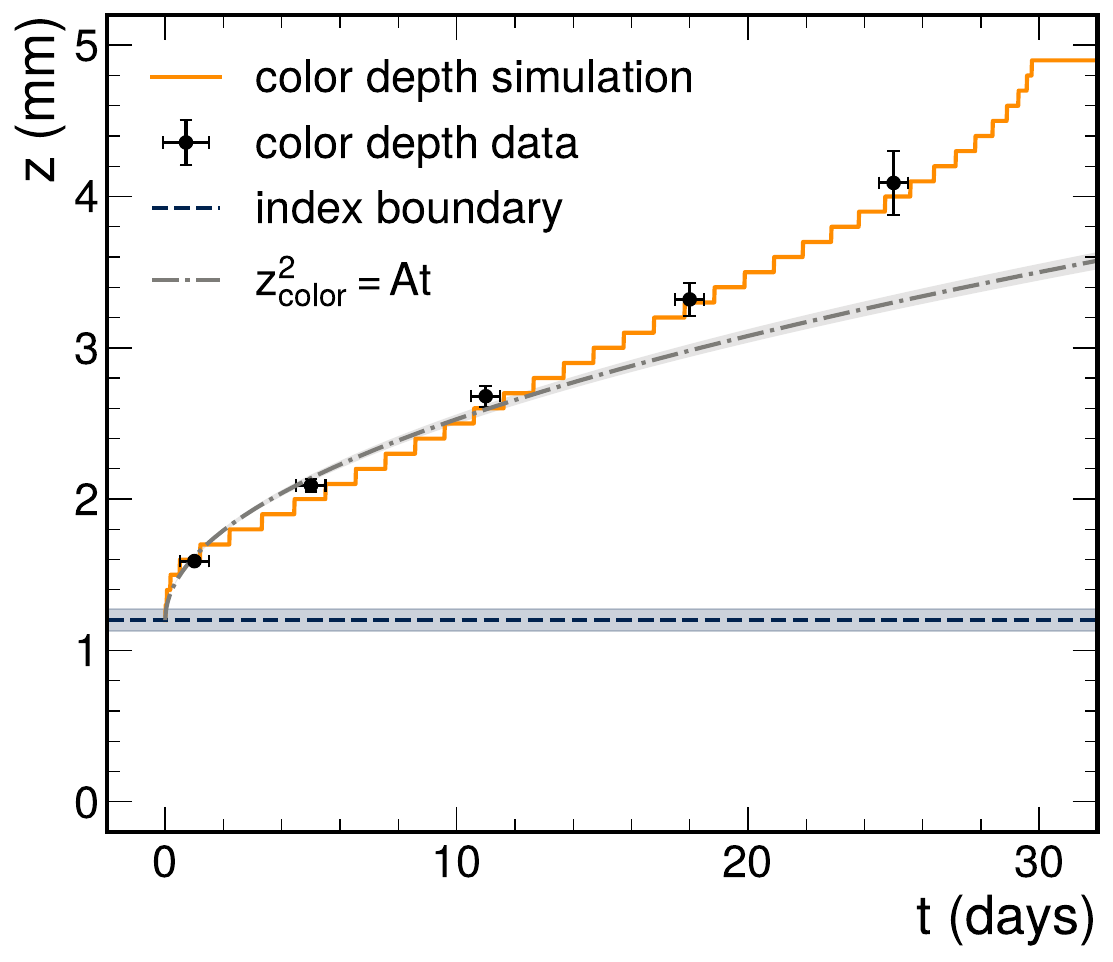}
    \caption{Measurements of the annealing process of an irradiated scintillator. The data points represent the depth of the colored region. The finite differences simulation of the colored region depth is shown with a yellow line. The depth of the refractive index boundary is shown with a dashed purple line. A fit of Eq.~\ref{eq:diff_progress} to the annealing data is shown as a green line.}
    \label{fig:annealing_depth}
\end{figure}

The annealing process was monitored closely and the depth of the colored region was measured periodically for a PS sample that was irradiated with $460\,$Gy/h for a total of $69\,$kGy. According to Ref.~\cite{Wick1991472}, the progression of the annealing is expected to follow the relationship
\begin{equation}\label{eq:diff_progress}
    z_{color}^2=At.
\end{equation}
The measurements of the colored region are shown in Fig.~\ref{fig:annealing_depth}
along with predictions of the finite differences simulation from Sec.~\ref{numSol1}, the position of the refractive index boundary, and a fit of Eq.~\ref{eq:diff_progress} to the data. The measurements evolve in accordance with the finite differences simulation whose input parameters have been tuned by minimizing the $\chi^2$ distance between the data and the prediction. Both the data and the simulation deviate away from Eq.~\ref{eq:diff_progress} after the initial part of the annealing process. Possible explanations for this might include the shape of the sample (the equation is derived for thin sheets) and the effect of the active radicals on oxygen diffusion. The fit result for the parameter in Eq.~\ref{eq:diff_progress} is $A_{PS}=7.351\pm0.015\cdot10^{-3}\,\mathrm{mm}^2/\mathrm{h}$, which is lower than the value $0.025\,\mathrm{mm}^2/\mathrm{h}$ from Ref.~\cite{Wick1991472}. This is expected since the fit is obviously underestimating the evolution rate.

\begin{figure}
    \centering
    \includegraphics[width=1.0\textwidth]{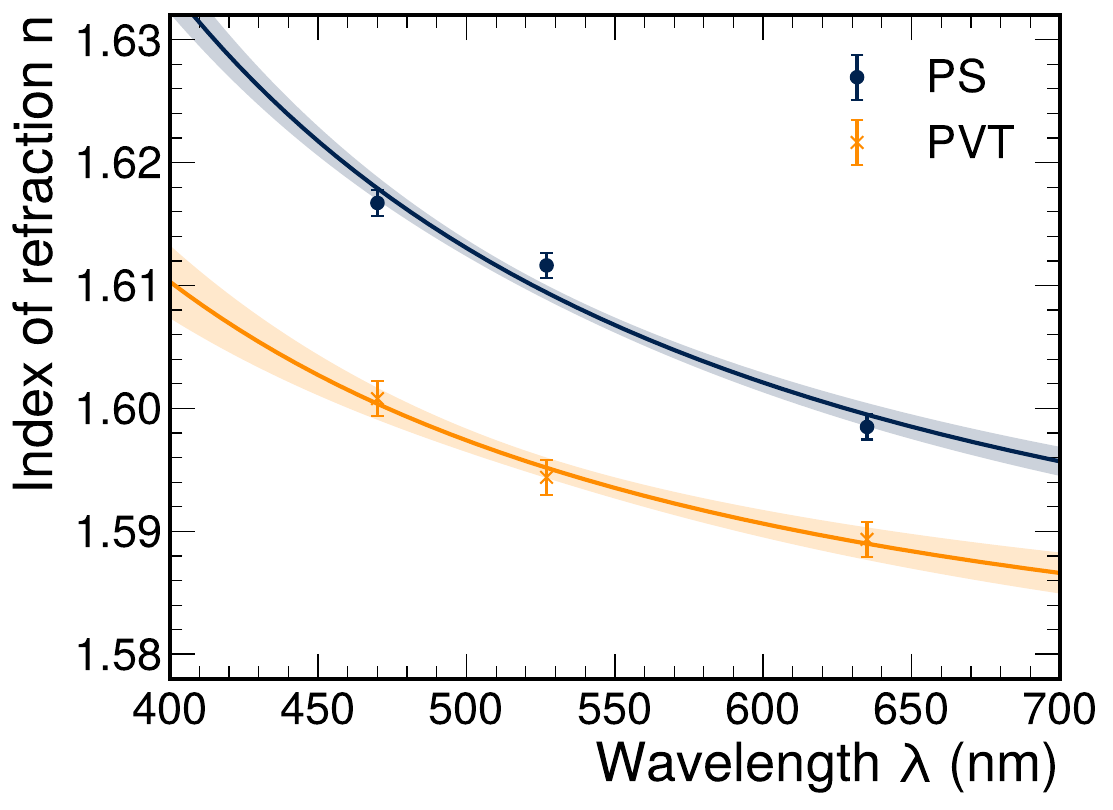}
    \caption{The refractive indices for unirradiated EJ-200 PS and PVT. The lines represent fits to the first-order Sellmeier equation. $1\sigma$ bands are shown around the fit lines.}
    \label{fig:unirr_index}
\end{figure}

The same method was used to measure the refractive indices of unirradiated EJ-200 PS and PVT scintillator samples. Fig.~\ref{fig:unirr_index}
shows the measured indices over the visible spectrum. Fits to the first-order Sellmeier equation are also included. The calculated Sellmeier coefficients for EJ-200 PS are $B^{\mathrm{PS}}_1=1.492\pm0.007$ and $C^{\mathrm{PS}}_1=1.71\pm0.13\cdot10^{-2}\,\mu\mathrm{m}^2$, and for EJ-200 PVT are $B^{\mathrm{PVT}}_1=1.483\pm0.011$ and $C^{\mathrm{PVT}}_1=1.10\pm0.19\cdot10^{-2}\,\mu\mathrm{m}^2$.

\begin{figure}[hbtp]
    \centering
    \includegraphics[width=0.49\textwidth]{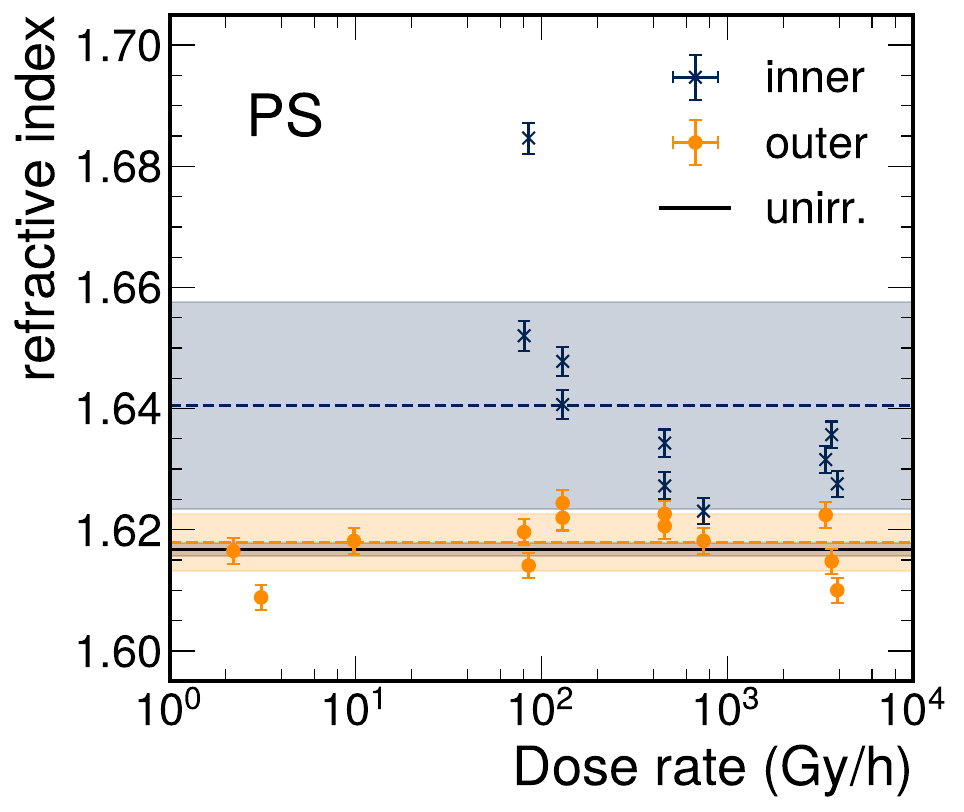}
    \includegraphics[width=0.49\textwidth]{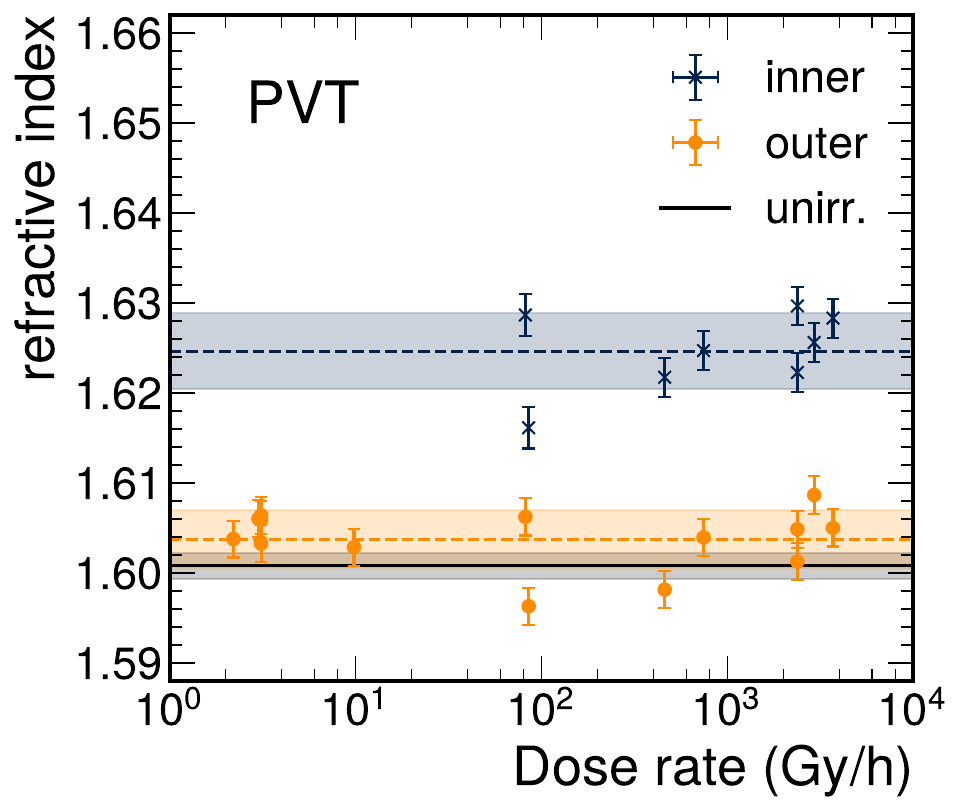}
    \caption{Measurements of the refractive index at $\lambda=470\,$nm after irradiation and annealing for EJ-200 PS (left plot) and PVT (right plot) for both the inner (blue) and outer (yellow) regions that are defined by the internal boundary. The average value for each region is shown with a dashed line and the standard deviation with a band around that line. The average of the measurements of the unirradiated samples is shown with a solid black line accompanied by a black-shaded area that represents the uncertainty.}
    \label{fig:refr_index}
\end{figure}

Scintillator samples of both substrates, that were irradiated over a very wide range of dose rates, were measured using the same technique. Fig.~\ref{fig:refr_index}
shows measurements of the index of refraction at the wavelength of $470\,$nm for irradiated scintillator samples of both PS (left plot) and PVT (right plot) substrates. For both materials, the index of the region inside the boundary is higher than the index in the outer region by approximately $\delta n\approx0.02$ on average. The samples at lower dose rates, that do not feature an internal boundary, have indices consistent with the values from the outer regions of the samples with boundaries. This is consistent with the hypothesis that these samples have been fully penetrated by oxygen during the irradiation. The values of the index for unirradiated samples are compatible with the indices of the outer regions within the experimental uncertainties. This suggests that non-oxidative processes, like radical crosslinking, are contributing more (at least collectively) to changes in the refractive index. Finally, there is no indication of dependence of the refractive index on the dose rate of the irradiation for the outer regions of both materials and the inner region of PVT. The refractive index of the inner region for the PS scintillators indicates a decreasing trend with increasing dose rate until it reaches a plateau slightly above the values of the outer region. The available data and their uncertainties prohibit the drawing of any conclusions regarding this trend, and therefore, more measurements with comprehensive coverage of the dose rate values are needed to evaluate this relationship.

\section{\label{sec:concl}Conclusions}

The effects of oxidation during irradiation on the optical properties of plastic scintillators were studied both theoretically and experimentally. Numerous models of oxygen diffusion in polymers, that can be solved numerically or analytically in limiting cases, exist in literature. Numerical solutions to the differential equations for two of those models show that the normalized oxygen reactivity can undergo a steep decrease from 1 to 0 at a certain depth within the material for a subset of the parameter space of the models. This prediction is in accordance with the optical observations made on irradiated polymeric scintillators that exhibit an optical boundary of sharp change in the refractive index, which separates the material into two regions. The depths at which these boundaries occur were measured and compared with the dose rates of the irradiations. The data show a dependence with the inverse of the square root of the dose rate, which is in agreement with the predictions of the models. Measurements of the index of refraction of the samples using a photography-based technique reveal that the internal regions defined by those boundaries have slightly elevated index values. The refractive indices of unirradiated samples agree with the values of the outer regions of the irradiated samples. The refractive indices do not depend on the dose rate of the irradiation with one exception, the inner region for PS for which a descending trend with increasing dose rate is observed. More studies are needed to clarify the validity and the nature of this trend.

\section*{Acknowledgments}
The authors would like to thank Chuck Hurlbut of Eljen Technology company for supplying many of the rods and for advice, and the staff at the Goddard Space Flight Center, the GIF++, and the National Institute of Standards and Technology for assistance with the irradiations. This work was supported in part by U.S. Department of Energy Grant DESC0010072.

\bibliography{main}

\end{document}